\documentclass[prd,twocolumn,a4paper,superscriptaddress,floatfix]{revtex4}
\usepackage{graphicx}
\usepackage{bbm}
\usepackage{amssymb}
\usepackage{calligra}
\usepackage{lipsum}
\usepackage[T1]{fontenc}
\usepackage[caption=false]{subfig}
\usepackage{float}
\DeclareFontShape{T1}{calligra}{m}{n}{<->s*[2.5]callig15}{}
\DeclareMathAlphabet{\mathcalligra}{T1}{calligra}{m}{n}
\DeclareMathAlphabet{\mathpzc}{OT1}{pzc}{m}{it}

\begin{document}
\newcommand{\be}{\begin{equation}}
\newcommand{\ee}{\end{equation}}
\newcommand{\bq}{\begin{eqnarray}}
\newcommand{\eq}{\end{eqnarray}}
\newcommand{\bsq}{\begin{subequations}}
\newcommand{\esq}{\end{subequations}}
\newcommand{\bc}{\begin{center}}
\newcommand{\ec}{\end{center}}
\newcommand\lapp{\mathrel{\rlap{\lower4pt\hbox{\hskip1pt$\sim$}} \raise1pt\hbox{$<$}}}
\newcommand\gapp{\mathrel{\rlap{\lower4pt\hbox{\hskip1pt$\sim$}} \raise1pt\hbox{$>$}}}
\newcommand{\dpar}[2]{\frac{\partial #1}{\partial #2}}
\newcommand{\sdp}[2]{\frac{\partial ^2 #1}{\partial #2 ^2}}
\newcommand{\dtot}[2]{\frac{d #1}{d #2}}
\newcommand{\sdt}[2]{\frac{d ^2 #1}{d #2 ^2}}
\newcommand{\vv}[0]{{\bar v}}
\newcommand{\ave}[1]{\left< #1 \right>}

\title{Signature of inflation in the stochastic gravitational wave background generated by cosmic string networks}

\author{G. S. F. Guedes}
\email[Electronic address: ]{up201101776@fc.up.pt}
\affiliation{Departamento de F\'{\i}sica e Astronomia, Faculdade de Ci\^encias, Universidade do Porto, Rua do Campo Alegre 687, PT4169-007 Porto, Portugal}

\author{P. P. Avelino}
\email[Electronic address: ]{pedro.avelino@astro.up.pt}
\affiliation{Departamento de F\'{\i}sica e Astronomia, Faculdade de Ci\^encias, Universidade do Porto, Rua do Campo Alegre 687, PT4169-007 Porto, Portugal}
\affiliation{Instituto de Astrof\'{\i}sica e Ci\^encias do Espa{\c c}o, Universidade do Porto, CAUP, Rua das Estrelas, PT4150-762 Porto, Portugal}
\affiliation{Centro de Astrof\'{\i}sica da Universidade do Porto, Rua das Estrelas, PT4150-762 Porto, Portugal}

\author{L. Sousa}
\email[Electronic address: ]{Lara.Sousa@astro.up.pt}
\affiliation{Instituto de Astrof\'{\i}sica e Ci\^encias do Espa{\c c}o, Universidade do Porto, CAUP, Rua das Estrelas, PT4150-762 Porto, Portugal}
\affiliation{Centro de Astrof\'{\i}sica da Universidade do Porto, Rua das Estrelas, PT4150-762 Porto, Portugal}

\begin{abstract}
A cosmic string network created during an inflationary stage in the early Universe --- here defined as i-string network --- is expected to enter a transient stretching regime during inflation, in which its characteristic length is stretched to scales much larger than the Hubble radius, before attaining a standard evolution once the network re-enters the Hubble volume after inflation. During the stretching regime, the production of cosmic string loops and consequent emission of gravitational radiation are significantly suppressed. Here, we compute the power spectrum of the stochastic gravitational wave background generated by i-string networks using the velocity-dependent one scale model to describe the network dynamics, and demonstrate that this regime introduces a high-frequency signature on an otherwise standard spectrum of the stochastic gravitational wave background generated by cosmic strings. We argue that, if observed by current or forthcoming experiments, this signature would provide strong evidence for i-strings and, therefore, for (primordial) inflation. We also develop a simple single-parameter algorithm for the computation of the stochastic gravitational wave background generated by i-strings from that of a standard cosmic string network, which may be useful in the determination of the observational constraints to be obtained by current and forthcoming gravitational wave experiments.
\end{abstract} 
\maketitle

\section{Introduction}

The recently inaugurated era of gravitational wave astronomy \cite{Abbott:2016nmj,TheLIGOScientific:2017qsa} opens the possibility of observing the primordial universe directly through the study of a variety of early universe gravitational wave sources \cite{Caprini:2018mtu}. Networks of cosmic strings --- linelike topological defects whose production as remnants of symmetry-breaking phase transitions is predicted in several grand unified scenarios \cite{vilenkin2000cosmic} --- are one such source. Although the production of cosmic strings is expected to occur in the early universe, they are generally expected to survive throughout cosmological history and the study of their observational imprints may give us insight into the phase transition that originated them.

As cosmic strings interact, they are expected to form closed loops that detach from the long string network. These loops then oscillate under the effect of their tension, and lose their energy in the form of gravitational waves. Since the production of these loops is predicted to occur copiously throughout the evolution of the Universe, they are expected to give rise to a characteristic Stochastic Gravitational Wave Background (SGWB) \cite{Vilenkin:1981bx,Hogan:1984is,Vachaspati:1984gt} that may be probed with a variety of gravitational wave experiments. There are then prospects of either detecting such a SGWB produced by cosmic strings, or of a significant tightening of current constraints on the string tension (and other model parameters) in the near future.

The cosmic-string-forming phase transition is often assumed to happen after primordial inflation is brought to an end. However, several models predict the formation of cosmic string networks during or towards the end of the inflationary stage \cite{PhysRevD.29.1870,Vishniac:1986sk,KOFMAN1987555,Yokoyama:1989pa,Freese:1995vp,Linde:1991km,Linde:1993cn,Jeannerot:2003qv,Rocher:2004my,1988PhLB..212..273Y}. In the braneworld realization of Superstring theory, inflation also often results in the production of cosmic superstrings \cite{Sarangi:2002yt,Jones:2003da,Binetruy:2004hh,Davis:2008dj} --- fundamental strings and 1-dimensional D-branes that may grow to macroscopic sizes and play the cosmological role of cosmic strings. In this paper, we shall study the SGWB generated by cosmic string networks produced during an inflationary stage in the early universe --- which for simplicity we shall refer to as i-strings. The accelerated expansion of the background is expected to rapidly stretch the i-strings, so that they become frozen in comoving coordinates, with a characteristic length significantly larger than the Hubble radius, at the end of inflation. Such a stretching regime cannot be maintained indefinitely after the inflationary stage ends. As a matter of fact, i-strings will eventually re-enter our Hubble sphere and the network will approach the evolution of a standard cosmic string network (produced after inflation). The effect of the phase of accelerated expansion is then to delay the attainment of the standard evolution: the more the network is stretched, the longer this delay is.

This feature of i-string networks --- the fact that they may only re-enter the horizon late in cosmological history --- has been generating some interest in the literature \cite{Kamada:2012ag,Kamada:2014qta,Ringeval:2015ywa} since these networks may evade the observational constraints on cosmic string tension that result from the Cosmic Microwave Background and the Pulsar timing array data. Note, however, that in these studies the authors only consider networks that become cosmologically relevant in the recent past. In the present paper, we shall consider the SGWB generated by i-string networks throughout their complete evolution. Since the production of cosmic string loops and the emission of gravitational waves are significantly suppressed for frozen networks, we shall show that the delay in the attainment of the standard evolution observed in i-string networks gives rise to a signature in the SGWB that may be observed in forthcoming gravitational wave experiments. Here we characterize this signature and argue that, if observed with upcoming experiments, it would provide unequivocal evidence for i-strings and, therefore, for an inflationary phase in the early Universe.

This paper is organized as follows. In Sec. II, the cosmological evolution of i-strings networks is described using the Velocity-dependent One Scale (VOS) model. In Sec. III, the emission of gravitational waves by cosmic string loops is characterized, and the method for the computation of the SGWB power spectrum is presented. In Sec. IV, the SGWB spectrum generated by i-strings is computed for a wide range of parameters. The signature of i-string networks is quantified and characterized as specific of a network that experienced an inflationary stage. In Sec. V, we develop a simple one-parameter algorithm for the quick computation of SGWB spectra generated by i-strings from SGWB spectra generated by standard networks created in the early Universe after inflation. We then conclude in Sec VI.

\section{Cosmological evolution of cosmic string networks created during inflation\label{vos}}

The Velocity-dependent One-scale (VOS)  \cite{Martins:1996jp,Martins:2000cs} model describes the evolution of a statistically homogeneous and isotropic cosmic string network  through the characterization of the evolution of the root-mean-square (rms) velocity of the network, $\bar{v}$, and its characteristic length, $L\equiv (\mu/\rho)^{1/2}$ (where $\rho$ is the average energy density of the network and $\mu$ the cosmic string tension). Considering the limit of infinitely thin strings, the evolution of these two quantities can be obtained by averaging the Nambu-Goto equations of motion \cite{Martins:1996jp,Martins:2000cs}:

 \bq
\label{eq:v}
\frac{d\bar{v}}{dt} & = & (1-\bar{v}^2)\left[\frac{k}{L}-\frac{\bar{v}}{\ell_d}\right] \,,\\
\label{eq:l}
\frac{dL}{dt} & = & LH+\frac{L\bar{v}^2}{2\ell_d} \,,
\eq
where $H=\dot{a}/a$  is the Hubble parameter, $a$ is the scale factor and a dot represents a derivative with respect to physical time. The damping lengthscale $\ell_d$ is defined as $\ell_d^{-1}=2H+\ell_f^{-1}$, where the first term accounts for the damping caused by the expansion of the Universe and the frictional length, $\ell_f$, encodes the frictional forces caused by interactions with other fields. Except when stated otherwise, for the remainder of this paper we shall assume that $\ell_f=\infty$, so that the only source of damping comes from the Hubble expansion.
Furthermore, $k$ is a velocity-dependent adimensional curvature parameter given approximately by \cite{Martins:2000cs}:
\be
\label{eq:k}
k(\bar{v})=\frac{2\sqrt{2}}{\pi}(1-\bar{v}^2)(1+2\sqrt{2}\bar{v}^3)\frac{1-8\bar{v}^6}{1+8\bar{v}^6}\, .
\ee

Interactions between cosmic strings may result in the formation of closed loops which detach from the network and oscillate, decaying radiatively. The energy density which is lost by the long string network as a result of the production of loops can be written as \cite{Kibble:1984hp}:

\be
\label{eq:loops}
\left.\frac{d\rho}{dt}\right|_{\rm loops}=\tilde{c}\bar{v}\frac{\rho}{L}\, ,
\ee
where $\tilde{c}$ is a phenomenological parameter that quantifies the efficiency of the loop-chopping mechanism. The value of $\tilde{c}=0.23\pm0.04$ has been shown to provide a good fit to cosmic string network simulations both in the radiation and matter eras \cite{Martins:2000cs}. Throughout this work we shall use this value for the parameter $\tilde{c}$.

The production and subsequent decay of cosmic string loops results in an additional energy loss term which needs to be taken into account in the VOS equations. This is done by adding the following term to the right hand side of the equation describing the evolution of the characteristic length [Eq. (\ref{eq:l})]:

\be
\label{eq:loss}
\left.\frac{dL}{dt}\right|_{\rm loops}=\frac{1}{2}\tilde{c}\bar{v}\, . 
\ee

Eqs. (\ref{eq:v}), (\ref{eq:l}) and (\ref{eq:loss}) constitute the VOS model and they allow for an accurate description of the cosmological evolution of cosmic string networks. The linear scaling regime \cite{PhysRevLett.60.257,PhysRevD.40.973,PhysRevLett.64.119,PhysRevD.45.R1000,Vincent:1996rb} --- during which $\bar{v}$ remains constant and $L=\xi t$ grows linearly with time ---  is an attractor solution of the VOS model in the case of a decelerating power-law expansion of the universe, where $a\propto t^{\beta}$, with $0<\beta<1$:

\be
\label{eq:scaling}
\xi=\sqrt{\frac{k(k+\tilde{c})}{4\beta(1-\beta)}} 
\quad\mathrm{and}\quad 
\bar{v}=\sqrt{\frac{k}{k+\tilde{c}}\frac{1-\beta}{\beta}} \,.
\ee
This regime is only attainable for a constant $\beta$ and, therefore, may only naturally arise in a realistic cosmological background deep into the matter or radiation epochs \cite{Avelino:2012qy}. Nevertheless, the existence of such a regime guarantees that the late-time cosmological evolution of cosmic string networks is the same for a large variety of initial conditions.

In this paper, we shall investigate the gravitational wave signatures generated by cosmic string networks produced during a primordial inflationary phase. These cosmic strings --- which, for simplicity, we shall refer to as i-strings --- are stretched and diluted as a result of the accelerated expansion of the cosmological background and, consequently, the characteristic length of the network grows very quickly. As a matter of fact, during an inflationary phase, one has \cite{Sousa:2011ew}

\bq
\label{eq:lpropa}
L & \propto & a \,,\\
\label{eq:vprop0}
\bar{v} & \propto & a^{-1-\frac{1}{\beta}}\to 0 \,.
\eq

As a result, at the end of inflation, $L$ may be significantly larger than the Hubble radius, with $LH\gg1$. A network of i-strings will then become frozen (in comoving coordinates) as a result of the rapid expansion and their subsequent evolution will be given initially by Eqs. (7) and (8). Note however that such a stretching regime is later brought to an end. For $0<\beta<1$ (as is the case in the radiation and matter eras), $LH\propto t^{\beta-1}$ decreases with physical time. Therefore, the characteristic length eventually becomes smaller than the Hubble radius --- the i-string network eventually thaws --- and the evolution of the network is no longer described by Eqs. (\ref{eq:lpropa}) and (\ref{eq:vprop0}). The stretching regime is then a transient one, and the network approaches the standard evolution afterwards.

\begin{figure}

\subfloat[]{

 \includegraphics[width=\linewidth]{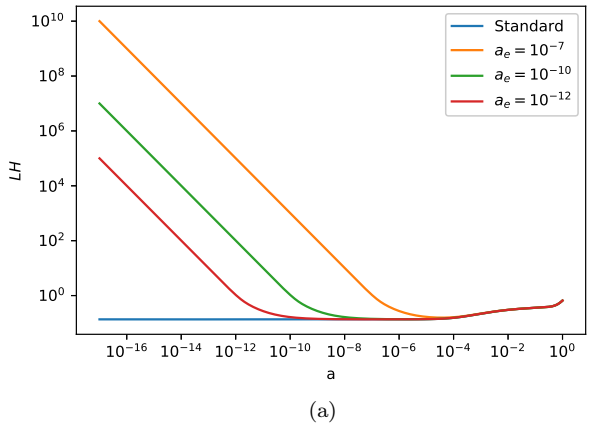}
\label{fig:lh}}

\subfloat[]{
  \includegraphics[width=\linewidth]{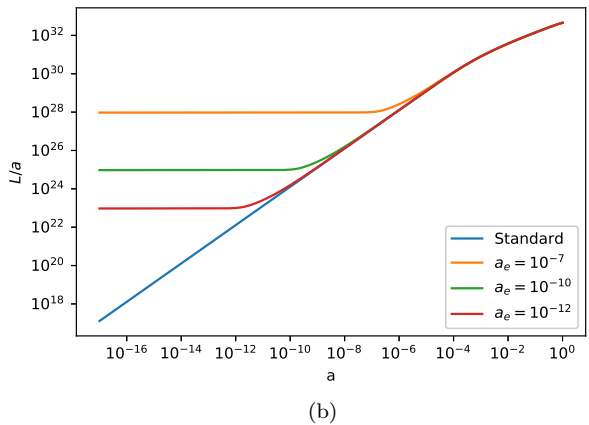}
  \label{fig:la}}

\subfloat[]{
  \includegraphics[width=\linewidth]{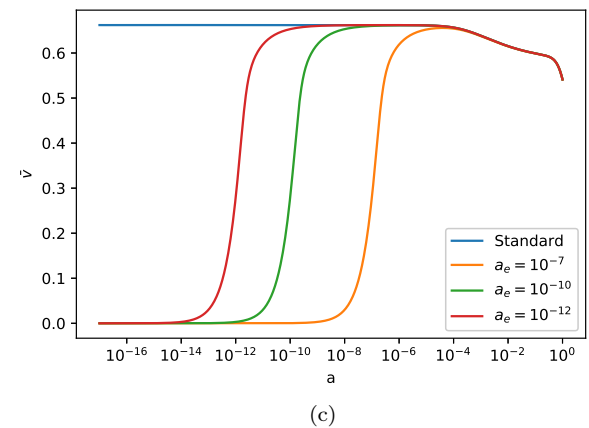}
  \label{fig:v}}

\caption{Evolution of $LH$ (a), $L/a$ (b) and $\bar{v}$ (c) with the scale factor $a$ for networks of cosmic strings with different entry times. The case labeled as standard represents a network for which $LH$ and $\bar{v}$ are set to their radiation-era scaling values as the initial conditions. The cosmological parameters used were $h=0.679$, $\Omega^0_\Lambda=0.694$ and $\Omega^0_rh^2=2.47\times10^{-5}$, in accordance with the Planck data \cite{Ade:2015xua}. The value of the scale factor at the present time is taken to be 1, $a_0=1$.}
\label{fig:subfigs}
\end{figure}

This behaviour is illustrated in Fig. \ref{fig:subfigs}, where the cosmological evolution of string networks which have $LH\gg1$ and $\bar{v}=0$ initially is plotted. For the calculations resulting in the evolutions depicted in Fig. \ref{fig:subfigs}, and all others in this work, the cosmological parameters used were $h=0.679$, $\Omega_\Lambda^0=0.694$ and $\Omega_r^0h^2 = 2.47\times 10^{-5}$, in accordance with the Planck data \cite{Ade:2015xua}. The value of the scale factor at present time, $a_0$, is taken to be equal to unity.

Let us denote the time for which $LH=1$ --- corresponding to the instant of time, after inflation, when the network re-enters the Hubble volume --- as the entry time. Fig. \ref{fig:lh} shows the cosmological evolution of $LH$ for networks with different entry times. Before $a_e$, the scale factor corresponding to the entry time, $LH$ follows a line with constant negative slope (as predicted in Eq. (7)) and afterwards evolves towards its scaling value, $\xi\beta$. The slope of the line before the entry time is the same for all cases since it is only influenced by $\beta$; therefore, a network created earlier in the inflationary era --- which experiences an accelerated expansion during a longer period of time and ends up with a larger characteristic length as a result --- re-enters the Hubble volume at a later time.

In Fig. \ref{fig:la}, the condition in Eq. (\ref{eq:lpropa}) becomes even more evident, with $L/a$ behaving as a constant before the entry time. For a network starting with a larger $L$, the transient stretching regime lasts longer and the i-string network thaws later in the cosmological history.

The same behaviour may be seen in Fig. \ref{fig:v}, which depicts the cosmological evolution of $\bar{v}$: as the scale factor approaches $a_e$, $\bar{v}$ (which was initially vanishing) quickly increases towards its scaling value as the network thaws. Following Eq. (\ref{eq:loops}), this implies that the network only starts producing a significant amount of loops near the entry time. 

As these figures show, an i-string network will eventually reach the evolution of standard cosmic string networks after a transient stretching regime, independently of the initial conditions. The effect of an early phase of accelerated expansion is then, in general, to delay the attainment of the standard evolution. Note however that, as these figures illustrate, the linear scaling regime shown in Eq. (\ref{eq:scaling}) cannot be established during the radiation-matter transition and that the matter era is not long enough for the network to re-establish scale-invariant evolution \cite{Avelino:2012qy}. In any case, even when the i-string network only re-enters the Hubble volume after the radiation-matter transition is triggered, it will still resume the evolution of standard cosmic string networks.

\section{The Stochastic Gravitational Wave Background generated by Cosmic String Networks}

The production of cosmic string loops as remnants of string interactions is expected to occur copiously throughout the cosmological evolution of cosmic string networks. Once these loops detach from the long string network they are, in general, expected to decay through the emission of gravitational radiation. As a matter of fact, these loops are predicted to emit Gravitational Waves (GWs) in a discrete set of frequencies determined by the length of the loops at the time of emission:

\be
\label{eq:frequencies}
f_j=\frac{2j}{l(t)}\,,
\ee
where $l(t)$ is the length of the cosmic string loop, $j=1,2,\cdots$ is the harmonic mode of emission and $f_j$ is the corresponding frequency.

Cosmic string loops are expected to emit gravitational radiation at a roughly constant rate

\be
\dtot{E}{t}=\Gamma G\mu ^2\,,
\ee
where $E=\mu l$ is the energy of the loops and $\Gamma \sim 65$ \cite{Vilenkin:1981bx,PhysRevD.42.2505} is a constant parameter which describes the efficiency of the GW emission mechanism. The length of the loop then decreases as a result of the emission of gravitational radiation, until they eventually radiate all their energy and disappear.

Cosmic string loops are thus predicted to generate a transient gravitational wave signal. However, since several loops are expected to exist at any given time in cosmic history, the superimposition of the bursts of GWs they emit in different directions is expected to give rise to a SGWB\cite{Vilenkin:1981bx,Hogan:1984is,Vachaspati:1984gt}. The amplitude of SGWB generated by cosmic string networks is often quantified using the energy density of GW, $\rho_{\rm GW}$, per logarithmic frequency interval (in units of the critical density $\rho_{\rm c}$):

\be
\Omega_{\rm GW}(f)=\frac{1}{\rho_{\rm c}}\dtot{\rho_{\rm GW}}{\log{f}}\,,
\ee
where $\rho_{\rm c}=3 H_0 ^2/(8\pi G)$ (the subscript `0' shall be used for the remainder of this paper to denote the value of the corresponding variable at the present time). This spectral density may be written as \cite{Sanidas:2012ee,Sousa:2013aaa}

\be
\Omega_{\rm GW}(f)=\sum_j^{n_s}\frac{j^{-q}}{\mathcal{E}}\Omega^j_{\rm GW}(f)\,,\label{highmodes}
\ee
where 
\be
\Omega_{\rm GW}^j(f)=\frac{16\pi}{3}\left(\frac{G\mu}{H_0}\right)^2\frac{\Gamma}{f a_0^5}\int_{t_i}^{t_0}j n\left(l_j(t'),t'\right)a^5(t')dt'\label{Omegagw}
\ee
is the contribution of the $j$-th harmonic mode of emission to the SGWB. Here, $t_i$ is the instant of time  associated to the start of loop production by the cosmic string network --- which is often assumed to be after friction becomes irrelevant to cosmic string dynamics, around $t_i\sim t_{pl}/(G\mu)^2$ \cite{vilenkin2000cosmic}, where $t_{pl}$ is the Planck time (we shall revisit this assumption in Sec. IV.B) ---, $n\left(l_j(t'),t'\right)dl$ is the number density of cosmic string loops with physical lengths between $l$ and $l+dl$ at the time $t$ and $l_j(t')=(2j/f)(a(t')/a_0)$ is the physical length that loops should have at a time $t'$ to radiate, in the $j$-th harmonic, GWs that have a frequency $f$ at the present time. Eq. (\ref{highmodes}) takes into account the fact that cosmic string loops emit GWs in different harmonic modes and the fact that gravitational backreaction is expected to damp higher frequency modes more efficiently than it does low-frequency modes \cite{Battye:1994qa,Battye:1997ji}. Therein

\be
\mathcal{E}=\sum_m^{n_s}m^{-q}\,
\ee
where $q$ is a parameter that depends on the shape of the loops, and $n_s$ is the number of harmonic modes that have been taken into consideration. It has been shown that $q=2$ for loops with one kink, while $q=4/3$ for loops with a cusp \cite{Vachaspati:1984gt}. Previous work \cite{Sanidas:2012ee} has shown that, in general, it is sufficient to consider modes up to $n_s=10^3$ or $n_s=10^5$ for loops with a kink or a cusp, respectively. In fact, for these values, the SGWB reaches a "saturation" and remains essentially unchanged by the inclusion of any higher order terms. For the remainder of this paper, we shall restrict ourselves to the fundamental mode of emission unless explicitly stated otherwise. However, note that, since

\be
\label{eq:modes}
\Omega_{\rm GW}^j(jf)=\Omega_{\rm GW}^1(f)\,,
\ee
one may easily construct $\Omega_{\rm GW}^j$ for any arbitrary mode of emission $j$, once the spectrum associated to the fundamental mode is computed.

\subsection{Loop distribution function}

As Eq. (\ref{Omegagw}) highlights, the loop distribution function, $n\left(l_j(t'),t'\right)$, is the pivotal quantity one has to characterize in order to compute the SGWB generated by cosmic string networks. Therefore, one needs to accurately estimate the size and number of cosmic string loops that exist at any instant in cosmic history in order to characterize their GW emission. However, numerical simulations of cosmic string networks have been inconclusive as to this regard. Although Nambu-Goto simulations --- in which cosmic strings are treated as infinitely thin and featureless objects --- have consistently shown that about $10\%$ of the energy lost by the network goes into the formation of large loops (with a size one to two orders of magnitude smaller than the horizon) \cite{Vanchurin:2005pa,Ringeval:2005kr,Olum:2006ix,Lorenz:2010sm,BlancoPillado:2011dq,Blanco-Pillado:2013qja}, field-theory simulations have found no evidence of a population of large loops \cite{Vincent:1997cx,JonesSmith:2007ne,Hindmarsh:2008dw,Figueroa:2012kw,Hindmarsh:2017qff}. As a matter of fact, in the latter, the main energy loss mechanism was observed to be the emission of scalar and gauge radiation instead of the production of loops --- a mechanism that is not taken into account in Nambu-Goto simulations and that is generally assumed to quickly become unimportant once the average distance between strings becomes significantly larger than their thickness. It is therefore currently unclear what is the dominant energy-loss mechanism in realistic cosmic string networks and what is the distribution of loops that are produced throughout their evolution.

Given this uncertainty, we shall take an alternative approach and make use of semi-analytical models to construct the loop distribution function of the cosmic string network. In this approach --- introduced in \cite{Caldwell:1991jj} and later extended \cite{Sanidas:2012ee,Sousa:2013aaa} --- the size of loops is treated as a free parameter and one has enough plasticity to probe a large variety of cosmic string scenarios. Here, we shall use the model introduced in \cite{Sousa:2013aaa} --- which is based on the VOS model to describe the cosmic string dynamics --- since it does not rely on assumptions of scale-invariant evolution and, therefore, it allows for an accurate computation of the loop distribution function during the radiation-matter transition. Note that therein, the authors found that an accurate modeling of cosmic string dynamics during this transition has a significant impact on the amplitude and broadness of the peak of the spectra and it is, thus, pivotal to make accurate predictions of the SGWB spectrum generated by cosmic string networks. 

Let us then assume that loops are created with a size that is a fixed fraction of the characteristic length of the network at the time of birth ($t_b$)

\be
l_b=\alpha L(t_b)\,,
\ee
where $\alpha$ is a constant parameter. Although one does not realistically expect all loops to be created at exactly the same size --- instead the distribution of the sizes of loops formed at the time $t_b$ is expected to peak around $l_b$ --- if the width of the distribution is not very large, this should be a good approximation (see Ref. \cite{Sanidas:2012ee} for a discussion of the effect of this assumption). In this approach, the number density of loops created as a function of time, $n_c$, may be estimated using the VOS model

\be
\dtot{n_c}{t}=\frac{1}{\mu\alpha L}\left.\dtot{\rho}{t}\right|_{\rm loops}=\frac{\tilde c}{\alpha}\frac{\bar v}{L^4}\,.\label{loopscreated}
\ee

Given the number density of loops that are created throughout cosmic history, the loops distribution function may easily be constructed for any $l$ and $t$. After formation, the size of loops decreases roughly linearly with time:

\be
\label{eq:looplength}
l(t)=\alpha L(t_b)-\Gamma G\mu (t-t_b)\,.
\ee
Therefore, $n\left(l_j(t'),t'\right)$ has contributions from all pre-existing loops that have a physical length $l_j(t')$ at time $t'$. Determining the times of creation ($t_b^i$) of the loops that contribute to a given frequency --- which one cannot do analytically if the networks are not in a linear scaling regime --- is therefore the essential step in this computation. Given these instants, the number density of loops is given by \cite{Sousa:2013aaa}

\be
n\left(l_j(t'),t'\right)  =  \sum_i \left\{ \frac{1}{\alpha \left. \frac{dL}{dt}\right|_{t=t_b^i}+\Gamma G\mu} \frac{\tilde c}{\alpha} \frac{\vv(t_b^i)}{L^4(t_b^i)}\left(\frac{a(t_b^i)}{a(t')}\right)^3\right\}\,.
\label{loopdist}
\ee

\subsection{The small-loop regime}

Small loops, with a length $l \ll \Gamma G\mu t$, are expected to live significantly less than a Hubble time, $t_H=H^{-1}$. Therefore, as demonstrated in 
\cite{Sousa:2014gka}, they can be regarded as decaying effectively immediately on the cosmological time-scale. Note, however, that the energy of small loops is not radiated in a single frequency, since the frequency of GWs is expected to increase as the length decreases. As a matter of fact, it was shown in \cite{Sousa:2014gka} that the power radiated by a small loop follows approximately the distribution

\be
p(f)=\frac{f_{\rm min}}{f^2}\theta(f-f_{\rm min})\,,
\ee
where $\theta(f-f_{\rm min})=1$ for all $f>f_{\rm min}$ and vanishes for all other $f$, and $f_{\rm min}=(2j/l_b(t))(a(t)/a_0)$ is the minimum frequency of emission of a loop (a loop of initial size $l_b(t)$ radiates for all frequencies $f>f_{\rm min}$). Note that this expression is not only valid for loops that are created with small sizes at time of birth, with $\alpha\ll \Gamma G\mu$, but also describes the emission of GWs at the end stages of the life of all loops irrespective of their initial size. As a matter of fact, as a large loop radiates energy in the form of GWs and its length decreases, it will eventually enter a regime in which its physical length is significantly smaller than the gravitational backreaction scale, $l \ll \Gamma G\mu t$, and its decay is precipitated.

The authors of Ref. \cite{Sousa:2014gka} devised an alternative method to compute the SGWB spectrum generated by small loops --- i.e., loops that are in the small-loop regime throughout their lifetime. In this case, the amplitude of the SGWB may simply be computed as follows

\be
\Omega_{\rm GW}^j(f)=\frac{16\pi G}{3H_0^2}\frac{j}{fa_0^5}\int_{t_{\rm min}}^{t_0}\left.\dtot{\rho}{t}\right|_{\rm loops}\frac{a^5(t)}{\alpha L}dt\,,
\label{omegasmall}
\ee
where $t_{\rm min}$ is the time of creation of the loops that have $f_{\rm min}=f$. Note that, if $t_{\rm min}\le t_i$, the lower bound is simply $t_i$. For small loops, this method produces identical results to the method described earlier in this section, with the advantage of requiring significantly less computational time.

\subsection{The typical SGWB spectrum}

Although the precise shape and amplitude of the SGWB spectrum generated by cosmic string networks is affected by a large variety of macroscopic and microscopic parameters (see e.g. \cite{Sanidas:2012ee,Sousa:2013aaa}) --- cosmic string tension and their large-scale dynamics, the size and emission spectrum of loops ---, this spectrum has a typical shape. 

The comoving characteristic lengthscale, $L/a$, generally grows throughout cosmological history and so does the comoving length of the loops that are produced. As a result, the cosmic string network forms loops which emit GW with a dominant energy density contribution to progressively lower frequencies. Therefore, the main contribution to the high frequency portion of the spectrum comes from smaller loops that were created deep in the radiation era and that have decayed at high redshifts. The GWs emitted by these loops generate a plateau in the high frequency range, whose amplitude is determined by cosmic string tension and the size of loops. Note that this plateau is sensible to the background expansion history, and, as a result, the alteration of the number of degrees of freedom caused by the annihilation of massive particles deep in the radiation-era is expected to generate step-like signatures in this plateau \cite{Caldwell:1991jj} (as matter of fact, any deviation from the standard background evolution is expected to generate a distinct signature in the spectra). However, these signatures will not be taken into account in the present study.

Note also that, for very high frequencies (with $f \gg f_{\rm min}(t_i)$), the only contributions to the spectra will come from the end stages of the life of loops created throughout cosmological history. This is merely a consequence of the fact that the production of cosmic string loops has not been happening since the beginning of the universe, instead it is expected to start being significant at a time $t_i$. Therefore, there are no loops emitting dominantly at frequencies $f \gg f_{\rm min}(t_i)$. The loops contributing to this frequency range do so in the small-loop regime and, therefore, it is straightforward to show --- using Eq. (\ref{omegasmall}) and taking into account the fact that, for $f \gg f_{\rm min}(t_i)$, $t_{\rm min}=t_i$ --- that in this portion of the spectrum $\Omega_{\rm GW}h^2\propto f^{-1}$.

On the other hand, in the low frequency range of the spectrum, there is a prominent peak generated by larger loops that emit GWs after the radiation-matter transition is triggered. The shape of this peak --- its location, amplitude and broadness --- is dependent on the size and emission spectrum of loops: larger loops, which live longer and therefore emit gravitational waves for a larger period of time, give rise to broader peaks with a higher amplitude.

\section{Stochastic gravitational wave background generated by network of cosmic strings created during inflation}

Several works predict the production of cosmic strings during or towards the end of an inflationary era \cite{PhysRevD.29.1870,Vishniac:1986sk,KOFMAN1987555,Yokoyama:1989pa,Freese:1995vp,Linde:1991km,Linde:1993cn,Jeannerot:2003qv,Rocher:2004my}. In this section, we characterize in detail the SGWB spectra produced by i-strings.

\subsection{Signature in the SGWB spectrum produced by i-strings}

As studied in Section II, a network of i-strings is stretched due to the accelerated expansion of the cosmological background in an inflationary era in the early universe, generally resulting in $LH\gg1$ at the end of inflation. The evolution of such a network was explored and, in Fig. \ref{fig:subfigs}, it has been shown that it attains the standard evolution at a later time. The fact that $\overline{v}$ is nearly vanishing while $LH\gg1$ implies that the production of loops is suppressed until the characteristic length of the network becomes of the order of the Hubble radius. Instead of producing a significant amount of closed loops shortly after creation, the network only starts producing them after the entry time.

\begin{figure}
  \includegraphics[width=\linewidth]{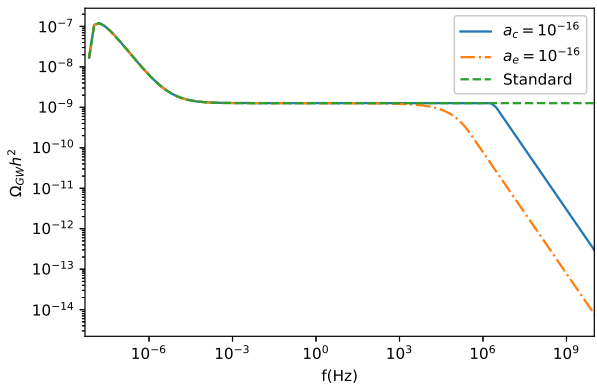}
\centering
  \caption{The SGWB spectra, $\Omega_{\rm GW}h^2(f)$, produced by a network of i-strings with $a_e=10^{-16}$ (orange dash-dotted line) and by a network of strings created at $a_c=10^{-16}$ (blue solid line). The standard spectrum is depicted in the dashed line. The spectra were calculated with $\alpha=10^{-9}$ and $G\mu=10^{-7}$.}
  \label{fig:compare}
\end{figure}

Since the general effect of experiencing an inflationary phase is to delay the onset of (significant) cosmic string loops production, one may then expect the SGWB spectrum generated by i-strings to be similar to the standard spectrum produced by a network of strings created near the entry time. This is illustrated in Fig. \ref{fig:compare} where the SGWB spectrum produced by i-strings with $a_e=10^{-16}$ is plotted along an artificial spectrum produced by a cosmic string network created in scaling at the time of entry, $a_c=10^{-16}$. As predicted, both spectra are very similar and coincide with the standard spectrum --- produced by a network of strings created in the early Universe after inflation --- in the low frequency range. Note, however, that there are significant differences in the approach to the standard spectrum: in the case of i-strings, after entry, the network approaches the scaling regime, while the artificial network created at the entry time is modeled to already be in the scaling regime upon formation. After the approach to the linear scaling regime, both networks produce loops at an equal rate, giving rise to the same spectra for lower frequencies as a result. At higher frequencies, both spectra present the $\Omega_{\rm GW}h^2\propto f^{-1}$ behaviour as a consequence of the lack of production of the loops which would contribute dominantly in this frequency range. As explained in Section III.C, this signature is produced by loops at the end stages of their lives, when they have already entered the small-loop regime. One can then conclude that the main signature of i-strings is to move this $\Omega_{\rm GW}h^2\propto f^{-1}$ behaviour to lower frequencies than those for which this signature would be observed in a standard SGWB spectrum.

\begin{figure}
  \includegraphics[width=\linewidth]{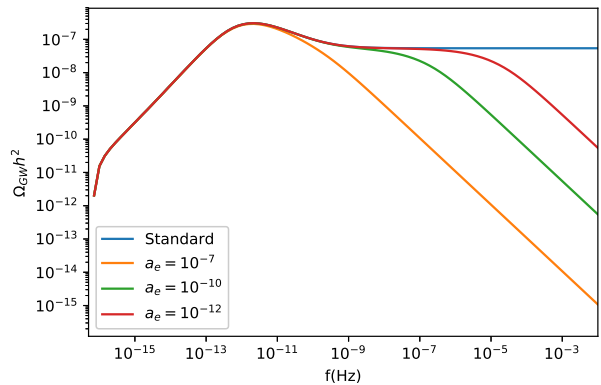}
\centering
  \caption{The SGWB spectra, $\Omega_{\rm GW}h^2(f)$, produced by networks with different entry times (the same as in Fig. \ref{fig:subfigs}). The spectra were calculated with $\alpha=0.1$ and $G\mu=10^{-7}$.}
  \label{fig:inf}
\end{figure}

Fig. \ref{fig:inf} shows the SGWB spectrum produced by networks of i-strings with different entry times and also a standard spectrum for comparison. The $\Omega_{\rm GW}h^2\propto f^{-1}$ signature is observed in all non-standard cases. Fig. \ref{fig:inf} shows that a later entry time results in a departure from the standard spectrum at lower frequencies. This is to be expected since a later entry time results in a longer delay of significant loop production, which means that the first loops created will contribute dominantly to lower frequencies.

\begin{figure}

\subfloat[]{

 \includegraphics[width=\linewidth]{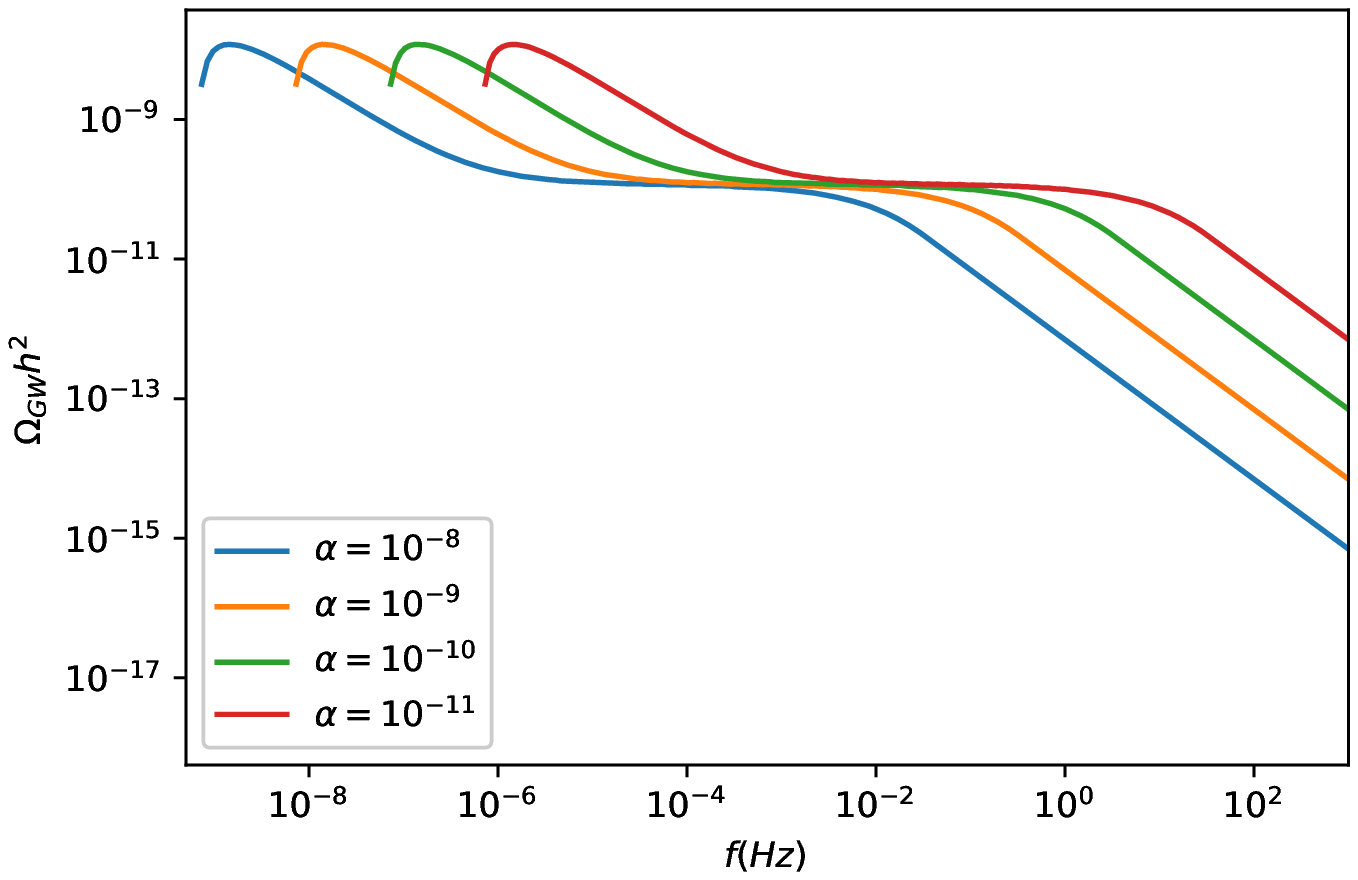}
\label{fig:difalfalarge}}

\subfloat[]{
  \includegraphics[width=\linewidth]{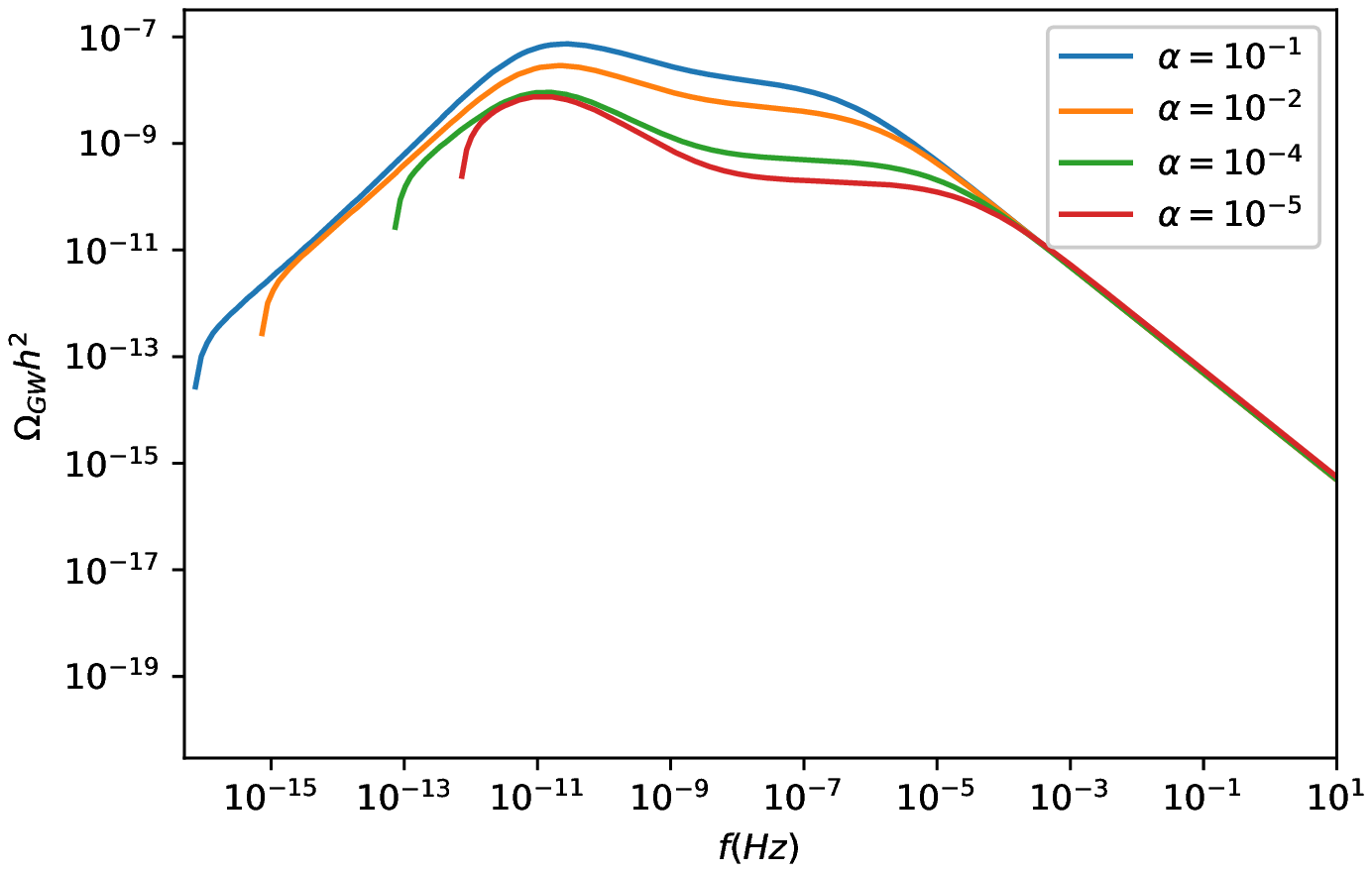}
  \label{fig:difalfasmall}}

\caption{The SGWB spectra, $\Omega_{\rm GW}h^2(f)$, produced by networks with different values of $\alpha$. In the top panel (a), cases where loops are considered large are shown while in the bottom panel (b) cases in the small-loop regime are plotted. The entry time is $a_e=10^{-10}$ and the spectra were calculated with $G\mu=10^{-8}$.}
\label{fig:subfigs2}
\end{figure}

\begin{figure}
  \includegraphics[width=\linewidth]{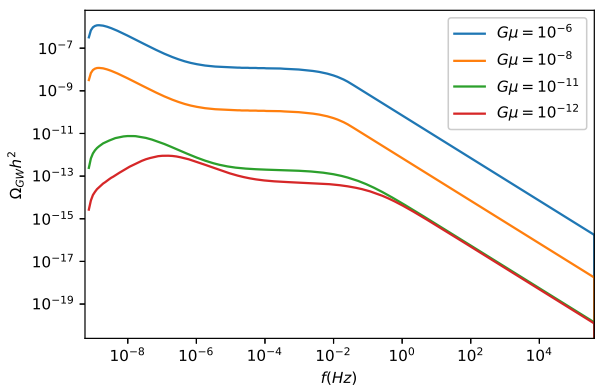}
\centering
  \caption{The SGWB spectra, $\Omega_{\rm GW}h^2(f)$, produced by networks with different values of $G\mu$. The entry time is $a_e=10^{-10}$ and the spectra were calculated with $\alpha=10^{-8}$.}
  \label{fig:difgmu}
\end{figure}

In Figs. \ref{fig:subfigs2} and \ref{fig:difgmu} the SGWB spectra generated by i-strings networks with $a_e=10^{-10}$ are plotted for different values of $\alpha$ and $G\mu$ respectively. In Fig. \ref{fig:subfigs2}, the cases for which the formed loops are considered large ($\alpha > \Gamma G\mu$)  are depicted in Fig. \ref{fig:difalfalarge}, while Fig. \ref{fig:difalfasmall} shows the cases in which the loops are considered small upon formation. The $\Omega_{\rm GW}h^2 \propto f^{-1}$ signature is produced by loops at the end of their lives, already in the small-loop regime regardless of their initial size, which explains why the behaviour is observed in every case for both figures. The dependence of $\alpha$ in the SGWB spectrum has been studied in \cite{Sanidas:2012ee,Sousa:2014gka}, which conclude, that in the regime of large loops, a larger $\alpha$ corresponds to a larger $\Omega_{\rm GW}$. Together with the fact that for a larger $\alpha$ the spectrum departs from the standard one at a lower frequency --- since $f_{min}$ of the first produced loops is lower for larger loops ---  this results in the $\Omega_{\rm GW}h^2 \propto f^{-1}$  signature almost coinciding for all cases. In the case of small loops this is not the case since a change in $\alpha$ results in a shift of the spectrum in the frequency axis which is observed in Fig. \ref{fig:difalfasmall}.

Fig. \ref{fig:difgmu} once again shows that the $\Omega_{\rm GW}h^2 \propto f^{-1}$ signature is present in every case. The distinction between large and small regimes is also present: the two spectra with lower energy density are the cases in which the formed loops are large and therefore the signature coincides in these cases.

This high-frequency $\Omega_{\rm GW}h^2\propto f^{-1}$ behaviour is therefore quite generic in the SGWB spectrum generated by i-string networks.  Since this signature implies a deficit of energy density at a specific range of frequencies, this fact has to be taken into account when analyzing observational data: the absence of detection of the predicted SGWB spectrum for specific parameters within this frequency range may mean that the network went through a period of accelerated expansion and not that it does not exist. As a matter of fact, as the figures presented in this section illustrate, if the i-string network only thaws late in cosmological history, the SGWB it generates may be undetectable even for relatively high tensions (see \cite{Kamada:2012ag} for a detailed characterization of the conditions under which i-strings may evade pulsar timing constraints). Note also that, even if a detection of a SGWB generated by cosmic string networks by pulsar timing experiment (which probe the low-frequency range) occurs in the near future, this spectra may not be observable with the more powerful LISA interferometer if the cosmic string network has experienced a sufficiently long inflationary stage.

\subsection{Specificity of the inflation signature}
We have shown that a $\Omega_{\rm GW}h^2 \propto f^{-1}$ signature in the SGWB spectrum produced by i-strings is imprinted by the first loops produced by the string network, those with the smallest comoving length. One may therefore ask whether or not the observation of a smooth $\Omega_{\rm GW}h^2 \propto f^{-1}$  high-frequency cut-off to an otherwise standard SGWB cosmic string spectrum should be taken as a signature of inflation. As we have seen in Sec. III.C, the (very) high-frequency portion of any cosmic string SGWB spectrum --- to which loops contribute at the end stages of their life --- is generally expected to exhibit a signature of the same kind. This question then reduces to whether these two signatures may be distinguished.

To answer this question let us consider the signature generated by a (standard) cosmic string network produced at a temperature $T_c$ in a string-forming phase transition occurring after inflation in the early Universe. In a weakly-coupled Higgs model \cite{vilenkin2000cosmic}
\be
G \mu \sim \left(\frac{T_c}{m_{pl}}\right)^2\,, \label{gmu}
\ee
and the corresponding time of string formation is given by 
\be
t_c \sim (g G \mu)^{-1} m_{pl}^{-1} \sim (g G \mu)^{-1} t_{pl}\,, \label{tc}
\ee
where $g=4 \pi (\pi {\mathcal N}/45)^{1/2}$. Here, we have used the Friedmann equation $H^2=8\pi G\rho/3$ and taken into account the fact that in the radiation era $H=(2t)^{-1}$ and $\rho=\pi^2 {\mathcal N} T^4/30$, where ${\mathcal N}$ is the number of effective relativistic degrees of freedom. We have also used fundamental units with $\hbar=c=k_B=1$, so that Planck mass is given by $m_{pl}=t_{pl}^{-1}=G^{-1/2}$.

The interaction of the strings with relativistic particles in a radiation fluid results in a frictional damping with a characteristic lengthscale \cite{Martins:1995tg,Martins:1996jp}
\be
\ell_f=\frac{\mu}{\sigma T^3}\,,
\ee
where $\sigma$ is a positive real number, associated to the number of species interacting non-minimally with the string (see also \cite{Sousa:2011ew,Sousa:2011iu} for a more general discussion of the role of friction on the dynamics of p-brane networks). Using Eqs. (\ref{gmu}) and (\ref{tc}), one may show that
\be
\ell_f \sim \sigma^{-1} g^{1/2} \left(t \, t_{pl}\right)^{1/2} \left(\frac{T_c}{T}\right)^2\,.
\ee
Since both $\sigma$ and $g$ are expected to be of order unity, $\ell_f (t_c)  \, H (t_c) \ll 1$ if $t_c \gg t_{pl}$. Hence, the initial evolution of the string network is expected to be friction dominated. As a matter of fact, the cosmic string network is expected almost immediately after being generated to attain a regime where $\ell_f H < LH < 1$ and $\bar{v} \sim k \ell_f / L$. If the average string density is low ($HL \gg {\tilde c} \bar{v}$), the network will experience a stretching regime --- during which it is conformally stretched, with $L \propto a$, $\bar{v} \propto \ell_f/a$ ---, before entering a transient Kibble phase with $H L \sim {\tilde c} \bar{v}$, and $\bar{v} \sim (k \ell_f H /{\tilde c})^{1/2}$, $L \sim (k {\tilde c} \ell_f/H)^{1/2}$. Both these phases precede the frictionless cosmic string evolution that we have described in Sec. II.

During the stretching phase the characteristic comoving lengthscale of the network $L/a$ remains roughly constant and the rms velocity of the strings is small. Hence, the generation of closed string loops by the network is limited, only becoming significant at the start of the Kibble phase. As a consequence, the first loops produced by the network, from the start of the stretching phase to the start of the Kibble phase, generate gravitational waves whose frequency measured at the present time is such that 
\be
f \gapp  t_c^{-1} \, \frac{a_c}{a_0}  \sim G \mu \, t_{pl}^{-1} \, \frac{a_c}{a_0}\,,
\ee
with 
\be
\frac{a_c}{a_0} \sim \frac{T_0}{T_c} \sim (G \mu)^{-1/2} \frac{T_0}{m_{pl}} \sim 2 \times 10^{-32} (G \mu)^{-1/2}\,,
\ee 
where $T_0 = 2.726 \, {\rm K}$ is the observed cosmic microwave background temperature \cite{Fixsen:2009ug}. Here, for simplicity, we neglected the fact that the entropy transfer to the photons from other species may slightly change this result if $T_c > 0.5 \, {\rm MeV}$, and we assumed that ${\mathcal N} = 3.36$ (only valid for $T_c < 0.5 \, {\rm MeV}$). These assumptions do not have any significant impact in the determination of the frequency range of the SGWB produced by the first loops created by the network, even if $T_c > 0.5 \, {\rm MeV}$. Here we have also assumed that these loops are created with the largest possible size, given by $\alpha\sim 1$, since we are interested in computing a lower limit to the frequency at which the initial $\Omega_{\rm GW}h^2 \propto f^{-1}$ signature occurs. Loops with smaller sizes would contribute at even higher frequencies.

Hence, the $\Omega_{\rm GW}h^2 \propto f^{-1}$ cut-off associated to a cosmic string network produced at a string-forming phase transition occurring after inflation only appears at frequencies 
\be
f \gapp 4 \times 10^{11} (G \mu)^{1/2}  \, {\rm Hz}\,,
\ee
where we have taken into account that $t_{pl}=m_{pl}^{-1}=5 \times 10^{-44} \, {\rm s}=7 \times 10^{-33} \, {\rm K}^{-1}$. For pulsar-timing experiments --- which currently provide the most stringent limits on the cosmic string SGWB --- the $\Omega_{\rm GW}h^2 \propto f^{-1}$ signature generated by cosmic string networks created after inflation would be within their frequency range (roughly $10^{-9}-10^{-6}\,{\rm Hz}$) for $G\mu\lesssim 10^{-41}$. This is far beyond the current reach of these experiments and, therefore, such a signature cannot be detected. The same is also true for the LIGO interferometer, which would only be able to detect this signature for $G\mu\lesssim 10^{-16}$. Moreover, one should not even expect the upcoming LISA interferometer to detect such a signature in spite of the significant increase in sensitivity it is expected to bring: this signature would only fall into its sensitivity window ($\sim10^{-5}-1\,{\rm Hz}$) for $G\mu\lesssim 10^{-23}$, which is also beyond its expected reach. This implies that either the value of $G \mu$ is small and there is insufficient power for this part of the spectrum to be observed with current and upcoming gravitational wave experiments, or this $\Omega_{\rm GW}h^2 \propto f^{-1}$ signature will be outside the frequency range covered by these experiments. 

Furthermore, during the Kibble phase, the average energy density of the cosmic string network is always higher than in the linear scaling regime, and the loop production is therefore extremely efficient (cf. Eq. (\ref{eq:loops})). This would generate a transitional region of frequencies in the SGWB power spectrum, between the standard radiation era plateau and the high frequency $\Omega_{\rm GW}h^2 \propto f^{-1}$ part, in which the power is significantly higher than that produced by a standard cosmic string network in a linear scaling regime (and, therefore, to some sort of a secondary peak). This feature is, in general not predicted in computations of the SGWB since networks are, for simplicity, generally assumed to start creating a significant amount of loops after the friction-dominated regimes have ended and scale-invariant evolution is established.

Hence, we conclude that the observation with current or forthcoming experiments of a smooth transition between a standard SGWB cosmic string spectrum and a $\Omega_{\rm GW}h^2 \propto f^{-1}$ spectrum is indeed a signature of a cosmic stringnetwork generated during inflation, assuming a standard transition from inflation to the radiation dominated era (see Refs. \cite{Cui:2018rwi,Cui:2017ufi} for a discussion of the signatures of non-standard cosmologies on the SGWB generated by cosmic strings).

Although this signature is associated with inflation, we should note that it does not distinguish between different inflationary scenarios nor does it differentiate inflation from its proposed alternatives (e.g., ekpyrotic models, bouncing and string gas cosmologies, etc...). Strictly speaking, the $\Omega_{\rm GW}h^2 \propto f^{-1}$ signature should more precisely be seen as an evidence that the characteristic lengthscale of the network was larger than the Hubble radius in the past.

\section{Recipe for the construction of spectrum\label{recipe}}
\begin{figure}
  \includegraphics[width=\linewidth]{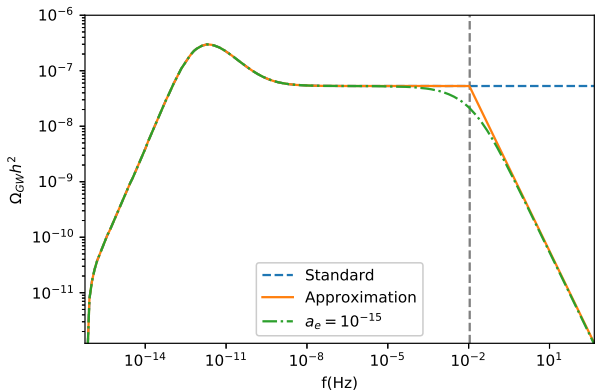}
\centering
  \caption{Approximation (solid line) of the SGWB spectrum, $\Omega_{\rm GW}h^2(f)$, following the recipe developed in this section. The dash-dotted line represents the SGWB spectrum produced by a network of i-strings, with $a_e=10^{-15}$ and the dashed line represents the standard SGWB spectrum. For low frequencies, these 3 spectra coincide. The vertical dashed line indicates the position of $f_{cut}$. The loop size parameter was set to $\alpha=0.1$ and $G\mu=10^{-7}$.}
  \label{fig:aprox}
\end{figure}

In this section we will provide a recipe to approximately construct the SGWB spectrum produced by a network of i-strings. We start with a standard spectrum and determine a specific frequency, $f_{cut}$, dividing it in two regions: for $f<f_{cut}$, the standard spectrum holds; while for $f>f_{cut}$, it becomes a line of constant slope -1 (with $\Omega_{\rm GW}h^2 \propto f^{-1}$), which intersects the standard spectrum at $f_{cut}$. This procedure is illustrated in Fig. \ref{fig:aprox}. The spectrum produced by a network of i-strings is represented by the dash-dotted line and our approximation --- which has a single parameter, $f_{cut}$ --- by the solid line.

From Fig. \ref{fig:aprox}, it should be noted that, while the approximation is accurate for the most part of the spectrum, the sudden transition from the standard spectrum to a straight line results in a slight overestimation of the amplitude of the spectrum for frequencies near $f_{cut}$. However, for the purpose of deriving observational constraints, this is not problematic: the resulting constraints will be safe, although a bit conservative.

The value of $f_{cut}$ is estimated through a numerical fit, to take into account the two effects responsible for the way the $\Omega_{\rm GW}h^2 \propto f^{-1}$ signature is added to an otherwise standard SGWB spectrum: the approximation to the linear scaling regime and the initial size of the loops --- since this signature is associated with the small loop behaviour.

\subsection{Case of small loops}
Studying loops which are created already in the small-loop regime it is possible to isolate the first effect, the approach to the scaling regime. As seen in the evolution of $\bar{v}$, shown in Fig. \ref{fig:v}, one expects the network to start producing a significant amount of  loops --- and therefore to start emitting a significant amount of gravitational waves --- near the entry time. However, by the entry time the network is not yet in a linear scaling regime (the value of $\bar{v}$ is approaching the scaling constant from below). This progression to the linear scaling regime results in a smooth transition from a standard spectrum to a region where $\Omega_{\rm GW}h^2 \propto f^{-1}$, as was seen in the previous section, in Fig. \ref{fig:inf}.

The value of $f_{cut}$ which takes into account this effect and produces the best approximation following our recipe, was numerically estimated to be the $f_{min}$ of loops created at $a_{e^*}=24\times a_e$, which corresponds to $LH=0.211$ --- as expected $LH$ is of order unity and close to the scaling value. From Eq. (\ref{eq:frequencies}) $f_{cut}$ can then be calculated as:

\be
\label{eq:cutsmall}
f_{cut}=f_{min}(a_{e^*})=\frac{9.48}{\alpha}H(a_{e^*})\frac{a_{e^*}}{a_0} \,.
\ee

\begin{figure}
  \includegraphics[width=\linewidth]{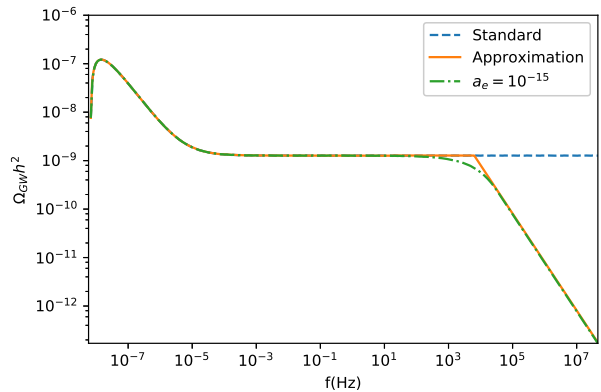}
\centering
  \caption{Approximation (solid line) of the SGWB spectrum, $\Omega_{\rm GW}h^2(f)$, following the recipe developed in this section. The dash-dotted line represents the SGWB spectrum produced by a network of i-strings, with $a_e=10^{-15}$ and the dashed line represents the standard SGWB spectrum. The loop size parameter was set to $\alpha=10^{-9}$ and $G\mu=10^{-7}$.}
  \label{fig:approxsmall}
\end{figure}

In Fig. \ref{fig:approxsmall} our approximation is plotted along a SGWB spectrum generated by i-strings for an entry time of $a_e=10^{-15}$, in the regime of small loops. The approximation accurately depicts the spectrum for a wide range of frequencies, except in a region near $f_{cut}$, where the overestimation that was mentioned in the beginning of this section can be seen in the sudden transition from the standard SGWB spectrum to a $\Omega_{\rm GW}h^2 \propto f^{-1}$ signature.

\subsection{Case of large loops}
For large loops, the fact that the loops created at $a_{e^*}$ only behave as small loops at a later time needs to be taken into account, since it is the small-loop behaviour that is responsible for the $\Omega_{\rm GW}h^2 \propto f^{-1}$  signature. To calculate this new $f_{cut}$ one has to define when does a loop behave as a small one. From section III, a loop can be considered to start behaving as a small one at the time $t_s$ given by:

\be
\label{eq:cutlarge}
l(t_s)<C\Gamma G \mu t_s\,,
\ee
where $C$ is a constant of order unity. For this particular recipe, the value of $C$ which originates the best approximation was found to be $C=0.6$. From this condition it follows that the time when a loop created at time $t_{e^*}$ (corresponding to the scale factor $a_{e^*}$) behaves as a small loop, $t_s$, can be calculated as

\be
\label{eq:timesmall}
t_s=\frac{1}{1.6\Gamma G\mu} \Bigg[\frac{0.211\alpha}{H(a_{e^*})}+ \Gamma G \mu t_{e^*}\Bigg]\,.
\ee

Knowing this time, $f_{cut}$ can be calculated following Eqs. (\ref{eq:frequencies}) and (\ref{eq:looplength}) as

\be
\label{eq:flarge}
f_{cut}=\frac{2}{\frac{0.211\alpha}{H(a_{e^*})}-\Gamma G\mu(t_s-t_{e^*})}\frac{a_s}{a_0}\,.
\ee

The calculation of $f_{cut}$ should be done following Eq. (\ref{eq:flarge}) if $\alpha > 2.84 \Gamma G \mu t_{e^*}H(a_{e^*})$. Otherwise, $t_s=t_{e^*}$ and Eq. (\ref{eq:flarge}) reduces to Eq. (\ref{eq:cutsmall}). 

Fig. \ref{fig:aprox} shows this approximation applied to a case in which the loops are large, with $\alpha=0.1$ and $G\mu=10^{-7}$. In this case we are clearly in the large-loop regime and the approximation accurately depicts the signature of  $\Omega_{\rm GW}h^2 \propto f^{-1}$. The overestimation is restricted to a region around $f_{cut}$.

\begin{figure}
  \includegraphics[width=\linewidth]{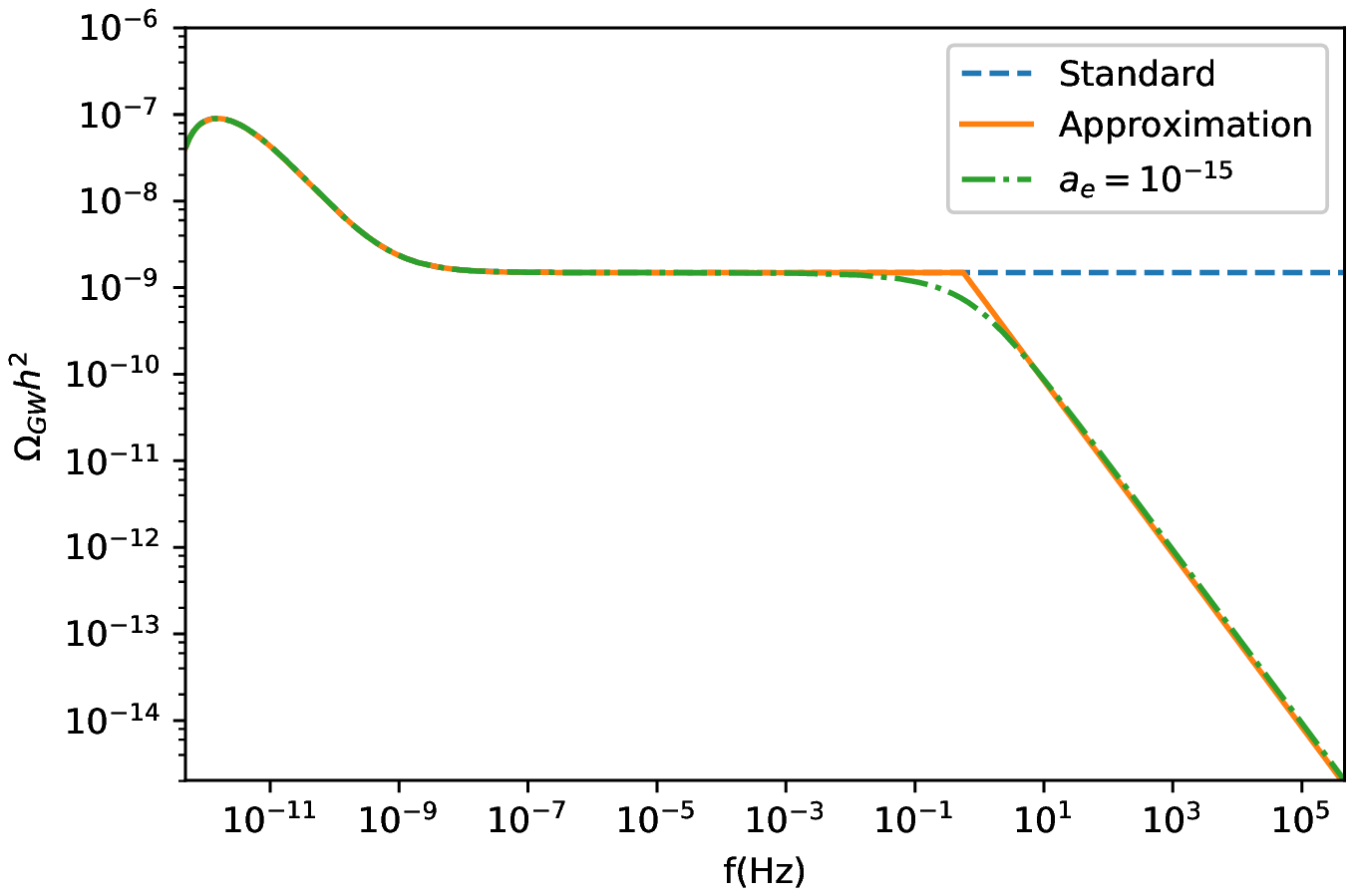}
\centering
  \caption{Approximation (solid line) of the SGWB spectrum, $\Omega_{\rm GW}h^2(f)$, following the recipe developed in this section. The dash-dotted line represents the SGWB spectrum produced by a network of i-strings, with $a_e=10^{-15}$ and the dashed line represents the standard SGWB spectrum. The loop size parameter was set to $\alpha=2\times10^{-5}$ and $G\mu=10^{-7}$.}
  \label{fig:approxtransition}
\end{figure}

A case near the transition between what is considered a small or a large loop for this recipe is shown in Fig. \ref{fig:approxtransition}. In this case $\alpha=10^{-9}$ and $G\mu=10^{-7}$. Despite working better for cases which are clearly in the small- or large-loop regimes, this approximation can still be used in the transition between these regimes --- with only minor deviations in the high-frequency range --- as evidenced by Fig. \ref{fig:approxtransition}. It is important to note that this transition was defined for this recipe in particular.

\begin{figure}
  \includegraphics[width=\linewidth]{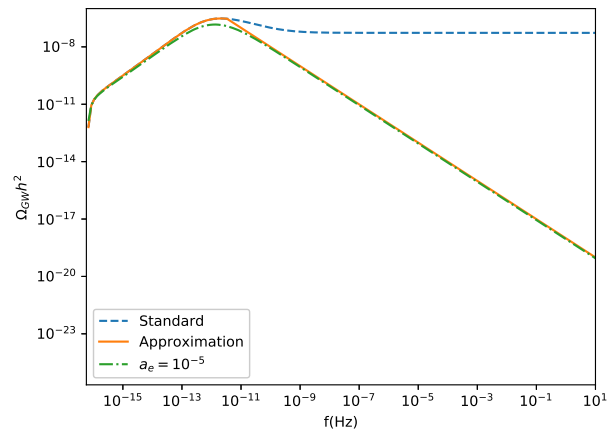}
\centering
  \caption{Approximation (solid line) of the SGWB spectrum, $\Omega_{\rm GW}h^2(f)$, following the recipe developed in this section. The dash-dotted line represents the SGWB spectrum produced by a network of i-strings, with $a_e=10^{-5}$ and the dashed line represents the standard SGWB spectrum. The loop size parameter was set to $\alpha=0.1$ and $G\mu=10^{-7}$.}
  \label{fig:approxmatter}
\end{figure}

This recipe provides a better fit to networks which reenter the Hubble radius during the radiation era, since the SGWB of standard networks is constant in this era and the network evolves in a scale-invariant manner in this epoch. The matter era may not be long enough for the network to reach scaling. However, this recipe can also be used when the reentry occurs during the matter epoch, as shown in Fig. \ref{fig:approxmatter}  --- where the departure from the standard spectrum happens near the peak. Given that $f_{cut}$ is not located in the plateau of spectrum (see Fig. \ref{fig:approxmatter}), the overestimation may be greater. However, even in this case the recipe may still be used to derive safe observational constraints.

From Eq. (\ref{eq:modes}) it becomes evident that it is straightforward to extend this approximation to any arbitrary harmonic mode of emission $j$. To do so one only needs to calculate $f_{cut}$ following the recipe described in this section and multiply it by $j$. Furthermore, following Eq. (\ref{highmodes}), one can construct the final spectrum including as many modes as intended. Nevertheless, such inclusion does not affect our main result, that is the $\Omega_{GW}h^2\propto f^{-1}$ signature associated with i-string networks.

\section{Conclusions}
In this paper we have computed the power spectrum of the SGWB generated by i-string networks --- networks of cosmic strings formed during inflation. Using the Velocity-dependent One Scale (VOS) model to describe the dynamics of the network, i-string networks were shown to enter a transient stretching regime during inflation in which the production of closed loops and the emission of gravitational waves are significantly suppressed, and the characteristic size of the network becomes much larger than the Hubble radius. Standard evolution is delayed until the characteristic length becomes again smaller or of the order of the Hubble radius. We have shown that this delay is responsible for a high-frequency signature of the form $\Omega_{\rm GW}h^2 \propto f^{-1}$ (the larger the delay is --- or, equivalently, the earlier the network is created during inflation --- the lower the frequency $f_{cut}$ at which this signature appears). We further argued that this signature, if observed by current or forthcoming experiments on an otherwise standard SGWB spectrum, would provide strong evidence for i-strings and, therefore, for (early) inflation.

In this work we also presented a simple single-parameter algorithm that allows for the construction of the SGWB spectrum produced by i-strings by using the the spectrum of standard networks as a starting point, without the need for a full recomputation. This algorithm provides an excellent approximation of the i-string SGWB spectrum for a wide range of parameters, including in the case of networks which only start producing a significant amount of gravitational waves in the matter-dominated era. This algorithm is expected to be useful in the determination of the observational constraints to be obtained by forthcoming gravitational wave experiments, or in the revision of current cosmic strings constraints (in particular, on the value of $G \mu$) to include the extra-parameter $f_{cut}$. Given that the i-string SGWB signature exhibits a deficit in power at high frequencies, it is possible that the lack of observation of the SGWB with current gravitational wave experiments might be associated to this deficit rather than a low value of $G \mu$.

Note that, despite the fact that this paper has focused on ordinary cosmic strings, the main results and conclusions of this paper also apply to the case of cosmic superstring networks. As a matter of fact, one shall also expect the SGWB generated by cosmic superstring networks that have experienced an inflationary stage to exhibit a signature of the form $\Omega_{\rm GW}h^2 \propto f^{-1}$ in the high-frequency range of the SGWB power spectrum (see \cite{Sousa:2016ggw} for a discussion of the SGWB generated by standard cosmic superstrings). Note also that a recipe such as that presented in Sec. \ref{recipe} may also in principle be used in this case.

\acknowledgments
L. S. was supported by Funda{\c c}\~ao para a Ci\^encia (FCT, Portugal) through the Grants No. SFRH/BPD/76324/2011 and No. CIAAUP-02/2018-PPD and through a work contract. This work was also supported by FCT through national funds (refª PTDC/FIS-PAR/31938/2017) and by FEDER - Fundo Europeu de Desenvolvimento Regional through COMPETE2020 - Programa Operacional Competitividade e Internacionaliza{\c c}\~ao (POCI) (refª POCI-01-0145-FEDER-031938). This paper benefited from the participation of P.P.A and L.S. on the European Cooperation in Science and Technology (COST) action CA15117 (CANTATA),supported by COST. Funding for this work was also provided by the FCT and by COMPETE2020 through these grants: UID/FIS/04434/2013 \& POCI-01-0145-FEDER-007672 and PTDC/FIS-PAR/31938/2017 \& POCI-01-0145-FEDER-031938.

\bibliography{biblio,istrings,friction}
\end{document}